\documentclass[journal]{IEEEtran}
\usepackage{graphicx}
\usepackage[thickspace,amssymb]{SIunits}
\pdfoutput=1
\begin{document}
\title{Properties of para-terphenyl as detector for $\alpha$, $\beta$ and $\gamma$ radiation.}
\author{%C.~Agodi$^{f}$, 
M.~Angelone$^a$, 
G.~Battistoni$^b$, 
F.~Bellini$^{c,d}$, 
V.~Bocci$^{d}$, 
F.~Collamati$^{c,d}$, %G.~Cuttone$^{f}$, 
E.~De~Lucia$^{e}$, %M.~De~Napoli$^{f}$, %A.~Di~Domenico$^{a,b}$, 
R.~Faccini$^{c,d}$, 
F.~Ferroni$^{c,d}$, 
S.~Fiore$^{f}$, %P.~Gauzzi$^{a,b}$, E.~Iarocci$^{d,e}$, 
M.~Marafini$^{d,g}$, 
D. ~Materazzo$^{h}$, 
I.~Mattei$^{e,i}$, %S.~Muraro$^g$, A.~Paoloni$^{c}$, 
S.~Morganti$^{d}$, 
V.~Patera$^{g,h}$, 
L.~Piersanti$^{e,h}$, 
M.~Pillon$^a$,
L. Recchia$^{d}$, 
A.~Russomando$^{c,j}$, %F.~Romano$^{e,f}$, 
A.~Sarti$^{e,h}$, 
A.~Sciubba$^{d,h}$, 
E.~Solfaroli~Camillocci$^{j}$, %E.~Vitale$^g$, 
C.~Voena$^{d}$ \\ % <-this % stops a space
$^a$ ENEA, Nuclear Fusion unit, Frascati, Italy;
$^b$ INFN Sezione di Milano, Milano, Italy;
$^c$ Dipartimento di Fisica, Sapienza Universit\`a di Roma, Roma, Italy; 
$^d$ INFN Sezione di Roma, Roma, Italy; 
$^e$ Laboratori Nazionali di Frascati dell'INFN, Frascati, Italy; 
$^f$ ENEA UTTMAT-IRR, Casaccia R.C., Roma, Italy;
$^g$ Museo Storico della Fisica e Centro Studi e Ricerche ``E.~Fermi'', Roma, Italy; 
$^h$ Dipartimento di Scienze di Base e Applicate per Ingegneria, Sapienza Universit\`a di Roma,  Roma, Italy;
$^i$ Dipartimento di Fisica, Universit\`a Roma Tre, Roma, Italy;
$^j$ Center for Life Nano Science@Sapienza, Istituto Italiano di Tecnologia, Roma, Italy.}%
%%\author{A.N. Author and A.N. Other\thanks
%%                 {On leave from another institue somewhere.}}
%%\institute{Institute name in English, Town, Country}
%\thanks{Manuscript received April 27, 2013.} %(Write the date on which you submitted your paper for review.) This work was supported in part by the U.S. Department of Commerce under Grant No. BS123456 (sponsor acknowledgment goes here).}% <-this % stops a space
\maketitle
\thispagestyle{empty}

\begin{abstract}
Organic scintillators are often chosen as radiation detectors for their fast decay time and their low Z, while inorganic ones are used when high light yields are required. In this paper we show  that a para-terphenyl based detector has a blend of properties of the two categories that can be optimal for energy and position measurements of low energy charged particles.
On 0.1\% diphenylbutadiene doped para-terphenyl samples we measure a light yield $3.5\pm 0.2$ times larger than a typical organic scintillator (EJ-200), and a rejection power for 660 keV photons, with respect to electrons of the same energy, ranging between 3-11\%, depending on the signal threshold. We also measure a light attenuation length $\lambda=4.73\pm 0.06 \ \milli \meter$ and we demonstrate that, with the measurements performed in this paper, a simulation based on FLUKA can properly reproduce the measured spectra.
\end{abstract}

%%%%%%%%%%%%%%%%%%%%%%%%%%%%%%%%%%%%%%%%%%%%%%%%%%%%%%%%%%%%%%%%%%%%%%%%%%%%%%

%%%%%%%%%%%%%%%%%%%%%%%%%%%%%%%%%%%%%%%%%%%%%%%%%%%%%%%%%%%%%%%%%%%%%%%%%%%%%%
\section{Introduction}
\IEEEPARstart{O}{rganic} scintillators are among the most common detectors for $\beta$ particles: these materials are capable of stopping electrons and positrons of few MeV within thicknesses of O(\centi \meter). Their scintillation light also has short decay times and is peaked in the blue region (400-450\ \nano \meter), matching the Photomultiplier Tubes' (PMT) spectral sensitivity. 

Organic scintillators have Light Yields (LY) of few thousand photons per \mega \electronvolt\ of energy released. This value is lower than what can be obtained with inorganic scintillators, but availability in large sizes and lower costs make organic scintillators preferrable for beta particles' detectors.

Aiming at closing the gap between inorganic and organic scintillators, we studied the properties of para-terphenyl, or p-terphenyl, an organic material commonly used as a dopant for organic scintillators, as the main component of a scintillator for radiation detection. %It is to be noted that the novelty in the approach is that p-terphenyl is the main component of the crystals we studied, while up to now it was used in scintillators just as a dopant.
Crystals of this material have a LY that strongly depends on the doping: LY can be significantly increased at the cost of shortening the Light Attenuation Length ($\lambda$).

P-terphenyl (1,4-diphenylbenzene) is an aromatic hydrocarbon isomer, formed by three benzene rings in ortho position. Pure terphenyl is a white crystalline solid, insoluble in water.
Though polychlorinated terphenyls were used as heat storage and transfer agents, p-terphenyl is currently under investigation as a material to be used in opto-electronic devices, such as organic LED devices (OLEDs) \cite{sasabe2011} and currently used in laser dyes and sunscreen lotions. In particle physics it has been used as a wavelength shifter, exploiting its sensitivity to VUV radiation, to read out scintillation light from liquid Xenon. P-terphenyl has also been used as a doping component for liquid scintillators \cite{Akimov2012,Akimov2010}. 
We investigated the scintillation properties of poli-crystalline p-terphenyl samples doped by 0.1\% in mass of diphenylbutadiene, with respect to organic scintillators.

The innovative aspect of our study is that p-terphenyl is the main component and not the doping agent. Due to this,  the crystals we studied have short $\lambda$, due to reabsorption of scintillation light. A short light attenuation length could be a counter-indication for common uses of scintillators, where the scintillating material is also the transport mean for the scintillation light. However, different approaches could benefit from this property. In particular, deposition of scintillating material on a substrate, coupling to  Wavelength Shifting fibers or substrates, or direct coupling to photodetectors will not be affected by the short $\lambda$. Moreover, dual-side readout of a crystal with charge-balance method could lead to the localization of the energy deposit along the crystal. Finally, detection of low energy charged particles does not require thick scintillators and can therefore cope with small attenuation lengths.

The response of a low Z, high LY and short $\lambda$ scintillator to low energy electrons, photons and $\alpha$s is mostly an unexplored field reported in this paper. 
The measured LY and $\lambda$ parameters are also a crucial input to the simulation of any detector built with such materials. Simulation is particularly critical in this case because the short attenuation length makes the tracing of optical photons compulsory. The capability of the FLUKA Monte Carlo~\cite{FLUKA,fluka2} to reproduce the response of p-terphenyl detectors using the measured scintillator properties is also shown.
%{\bf{AGGIUSTARE ORDINE REFERENZE}}
%%%%%%%%%%%%%%%%%%%%%%%%%%%%%%%%%%%%%%%%%%%%%%%%%%%%%%%%%%%%%%%%%%%%%%%%%%%%%%

%%%%%%%%%%%%%%%%%%%%%%%%%%%%%%%%%%%%%%%%%%%%%%%%%%%%%%%%%%%%%%%%%%%%%%%%%%%%%%
\section{Experimental Setup}
\label{pterp}

The absolute measurement of the LY of a scintillator depends on the shape and polishing of the sample, on the wrapping and optical coupling to the photodetector, and on the photodetector response. For this reason 
we performed relative LY measurements by using reference samples of commercial  EJ-200 plastic scintillator  from Eljen~\cite{eljen}, whose LY is tabulated to be $\simeq$10000 photons/1 MeV e$^-$.
The EJ-200 and p-terphenyl samples we used were $32 \ \milli \meter$ diameter, $3$, $4$ and $5$\ \milli \meter\  height cylinders. P-terphenyl samples have been also combined base to base with BC 630 optical grease to increase the total thickness $T$ traversed by scintillation light.

\label{setup}

In order to study the performances of p-terphenyl as a detector for $\alpha$ particles, electrons and photons, we exposed the samples to a $^{241}$Am $\alpha$ source and a  $^{137}$Cs electron and $\gamma$ source.

To collect scintillation light, the samples were optically coupled to an Hamamatsu H10580 Photomultiplier (PMT), whose spectral sensitivity is peaked at 420\ \nano \meter. The typical Quantum Efficiency for this PMT is of 25\%.
%The samples were coupled to the PMT with optical grease. If not otherwise specified, the source was placed in contact with the scintillator in order to reduce the material traversed by the radiation. The overall setup was wrapped with aluminum.
One base of each sample was coupled to the PMT with optical grease, while the other surfaces were covered with aluminized mylar. If not otherwise specified, the source was placed in contact with the scintillator in order to reduce the material traversed by the radiation.

%%%%%%%%%%%%%%%%%%%%%%%%%%%%%%%%%%%%%%%%%%%%%%%%%%%%%%%%%%%%%%%%%%%%%%%%%%%%%%

%%%%%%%%%%%%%%%%%%%%%%%%%%%%%%%%%%%%%%%%%%%%%%%%%%%%%%%%%%%%%%%%%%%%%%%%%%%%%%

\section{Light Attenuation Length}
\label{LAL}

\begin{figure}[!h]
\begin{center}
\includegraphics [width = 0.45  \textwidth] {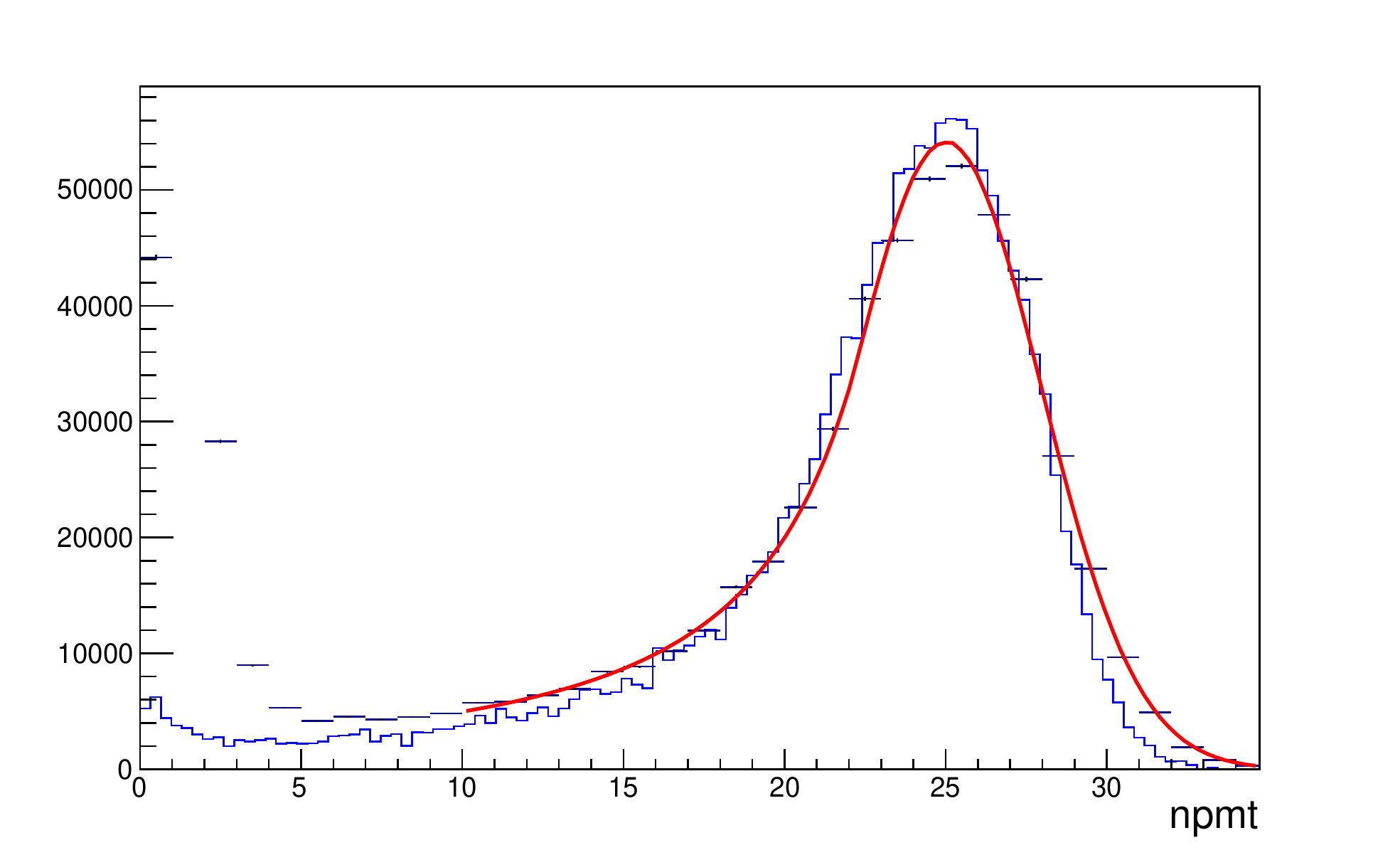}
\caption{Signal spectrum, in units of PMT charge, for the 8 mm p-terphenyl sample exposed to the $^{241}$Am $\alpha$ source (dots). The tuned MC spectrum (properly normalized and calibrated) is superimposed (solid line histogram). The superimposed fit is described in the text.}
\label{fig:alphaspectrum}
\end{center}
\end{figure}

\begin{figure}[bth]
\begin{center}
\includegraphics [width = 0.45  \textwidth] {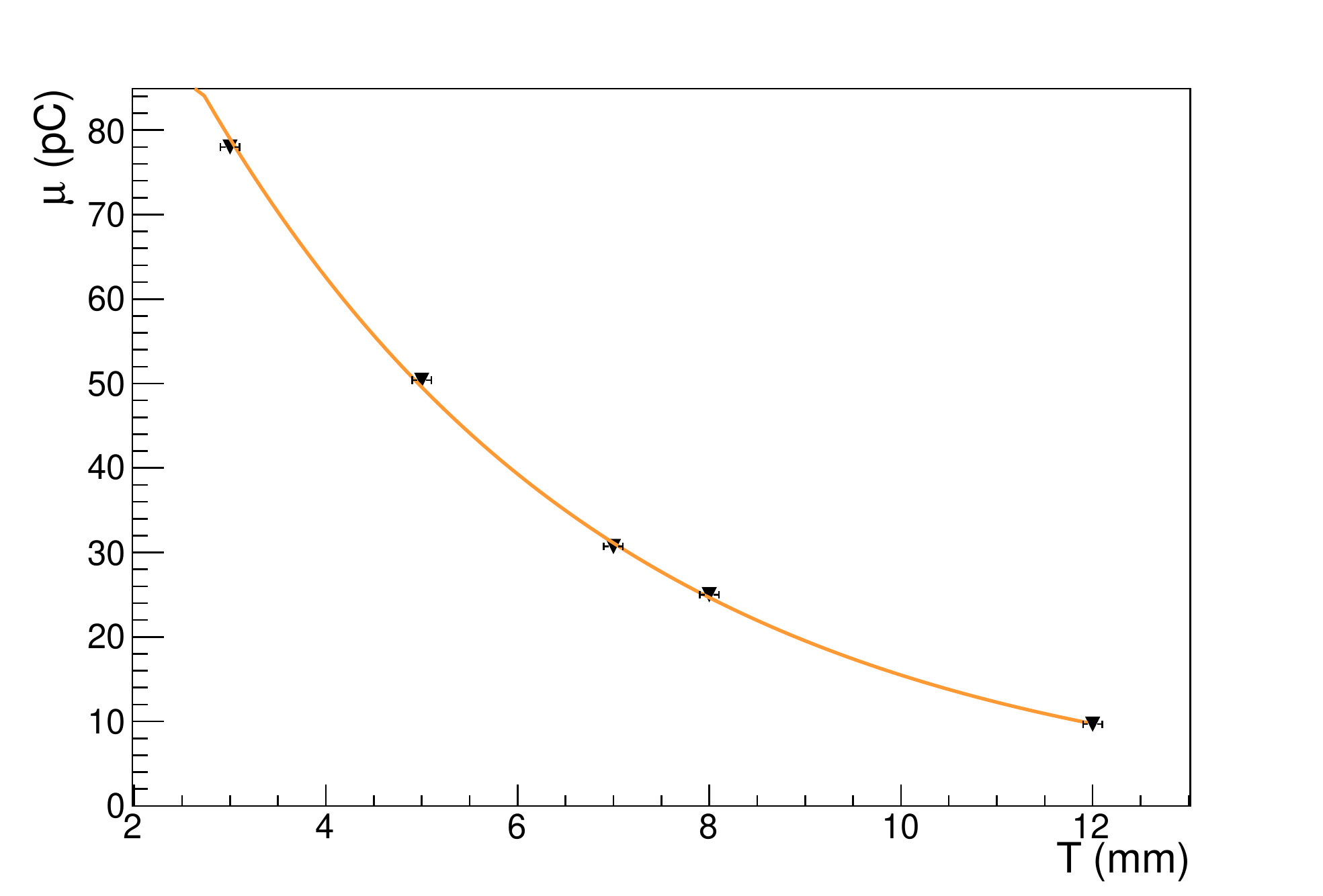}
\caption{Dependence of the mean signal from a $^{241}$Am $\alpha$ source on p-terphenyl samples as a function of their thickness. The superimposed exponential fit is described in the text. 
%Right: comparison between data and MC of the resolution as a function of the crystal depth. 
}
\label{fig:FITalpha}
\end{center}
\end{figure}

The first investigated property of p-terphenyl is the Light Attenuation Length ($\lambda$) since it affects all other measurements. In order to  measure it, a non penetrating radiation is needed since otherwise the measured quantity would be equal to the convolution between the penetration path distribution and the $\lambda$. We therefore exposed samples with different thickness $T$ to the non-penetrating radiation of $\alpha$-particles from a $^{241}$Am source, producing two indistinguishable $\alpha$ lines with mean energy $E_\alpha=5.48\ \mega \electronvolt$. The energy spectra for $\alpha$ particles in the p-terphenyl samples, in units of PMT integrated charge, have been fit with a {\it Crystal Ball} function
\footnote{The Crystal Ball function $CB(x|\mu,\sigma,a,n)$ is a gaussian with width $\sigma$ and mean $\mu$ for $(x-\mu)\ge -a\sigma$, and a power law $A(B-x)^n$ for  $(x-\mu)<-a\sigma$.} 
as shown in Fig.~\ref{fig:alphaspectrum}. The dependence of the measured  mean released energy ($\mu$) on the sample thickness (Fig.\ref{fig:FITalpha}) is fitted with an exponential function, and the resulting effective light attenuation length is $\lambda_m=4.28\pm 0.06$\ \milli \meter. The error includes the systematic error that is estimated by adding in quadrature a constant error contribution chosen in order to enforce the $\chi^2$ of the fit to equalize the number of degrees of freedom. 

Due to their short range, $\alpha$ particles release all their energy close to the surface of the samples. Scintillation light will traverse all the sample before reaching the PMT, so the collected light at the PMT photocathode will have travelled for the whole sample thickness, plus an additional length due to optical photon angular spread. The obtained attenuation length is therefore shorter than the true one.
In order to account for this effect, measurements with several thicknesses were simulated with FLUKA (see Sec.~\ref{MC})
and the resulting attenuation length of p-terphenyl is measured to be
\begin{equation}
\lambda=4.73\pm 0.06\ \milli \meter.
\end{equation}

%%%%%%%%%%%%%%%%%%%%%%%%%%%%%%%%%%%%%%%%%%%%%%%%%%%%%%%%%%%%%%%%%%%%%%%%%%%%%%

%%%%%%%%%%%%%%%%%%%%%%%%%%%%%%%%%%%%%%%%%%%%%%%%%%%%%%%%%%%%%%%%%%%%%%%%%%%%%%

\section{ $\alpha$ quenching factor and Electron Light Yield}
\label{LY2}

The short $\lambda$ of our p-terphenyl samples makes the measurement of the LY extremely difficult. The light yield should represent the number of emitted photons  per unit of released energy, but the quantity of practical interest for scintillation detectors is the number of photons reaching the photo-detector ($L$). Such number depends on the distance $l$ travelled by the light from the position of the energy release to the photo-detector, and on the scintillator $\lambda$: $L=LY\cdot e^{(-l/\lambda)}$. %The latter can be corrected for knowing the attenuation length, but a reasonable 
An independent measurement of LY thus requires the use of radiation %that releases the energy at a fixed depth, i.e. 
with a penetration range shorter than $\lambda$.

While $\alpha$s have the lowest penetration range, the quenching factor for their deposited energy implies the correspondent LY to be lower than the one for electrons. LY measurement through electrons  requires instead the use of monochromatic electrons with energy low enough to penetrate distances smaller than $\lambda$.

We found the electrons from $^{137}$Cs to match these requirements. $^{137}$Cs decays 94.7\% of the times in $^{137}$Ba$^*$ via $\beta$ decay, emitting an electron with maximum energy of $523\ \kilo \electronvolt$; $^{137}$Cs decays in the remainder of the cases into $^{137}$Ba via $\beta$ decay, with a maximum energy of $1176\ \kilo \electronvolt$. $^{137}$Ba$^*$ de-excites to its ground state by emitting a $662\ \kilo \electronvolt$ photon. The emitted photon has an 18\% probability to transfer its energy to one $^{137}$Ba electron by internal capture, yielding an electron with energy equal to the photon energy minus the escape energy. The electrons relevant to this study have an energy of $624\ \kilo \electronvolt$ (with a 7.8\% probability) and $656\ \kilo \electronvolt$ (with a 1.4\% probability)~\cite{CsSpectrum}. 

The same source produces also a $662\ \kilo \electronvolt$ $\gamma$, but unfortunately the low Z of the organic scintillators implies a very low detection efficiency. This aspect will be detailed in Sec.~\ref{phot}.

\begin{figure}[bth!]
\begin{center}
\includegraphics [width = 0.45  \textwidth] {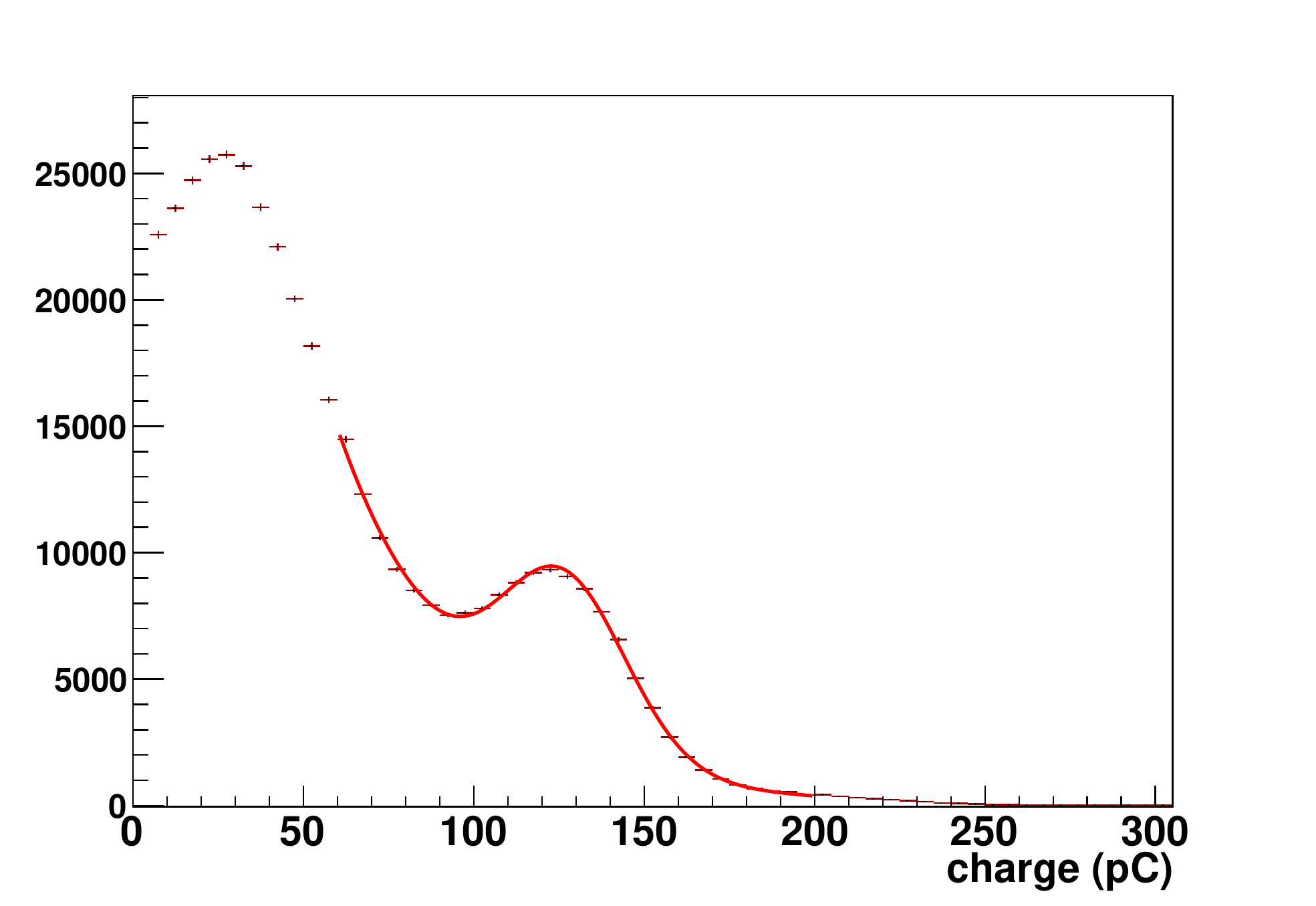}
\caption{Charge spectrum in the 4mm p-terphenyl samples irradiated by $^{137}$Cs. The superimposed fit is described in the text.}
\label{fig:betacs}
\end{center}
\end{figure}

Energy spectra were acquired with the three p-terphenyl samples  and combinations of them (see Sec.\ref{pterp}). 
Fig.~\ref{fig:betacs}  shows a typical charge spectrum with the fit to an exponential background and a gaussian signal superimposed. The $\mu$ and $\sigma$ parameters of the gaussian function will be referred to as mean signal and standard deviation, respectively.

\begin{figure}[bth!]
\begin{center}
\includegraphics [width = 0.45  \textwidth] {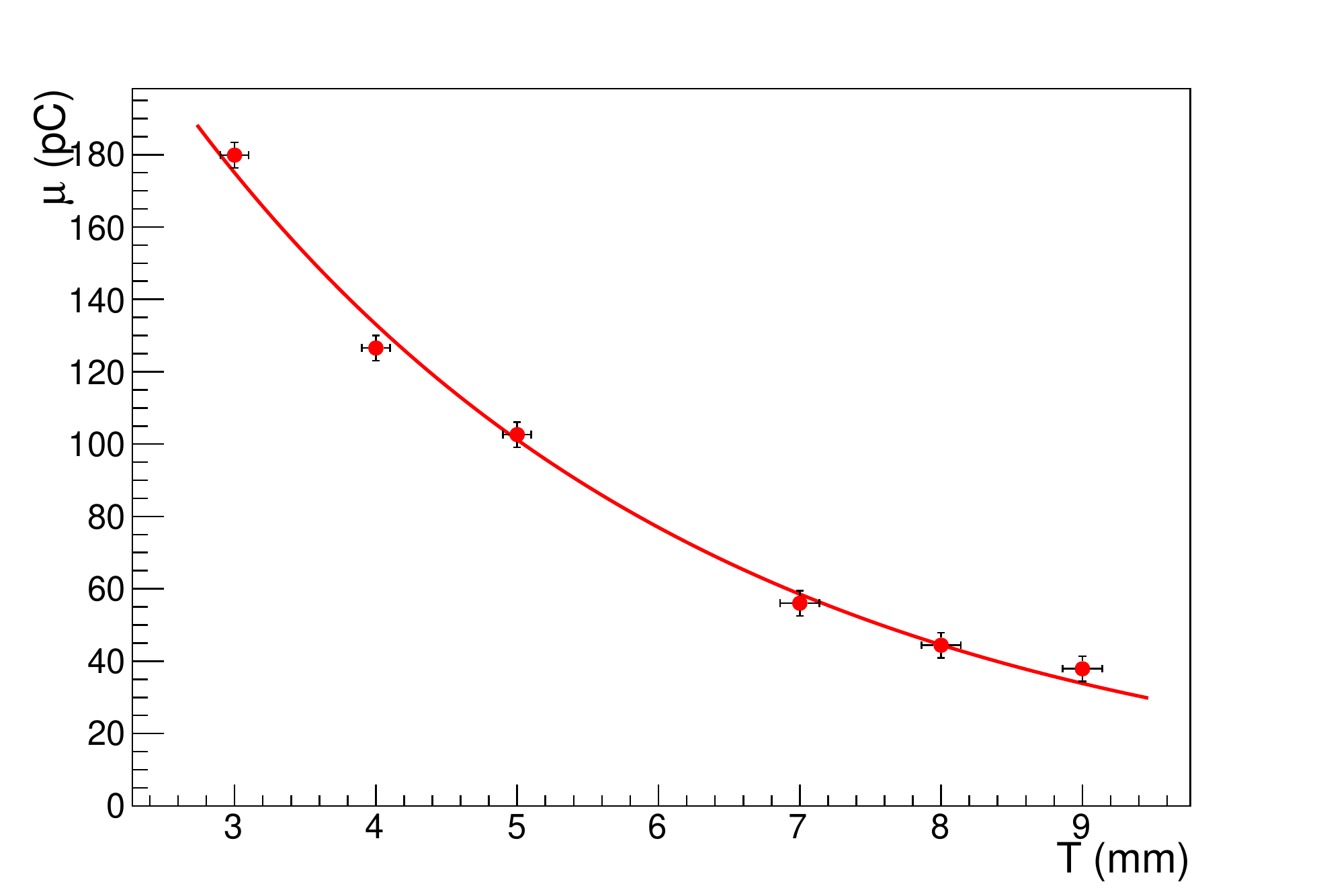}
\includegraphics [width = 0.45  \textwidth] {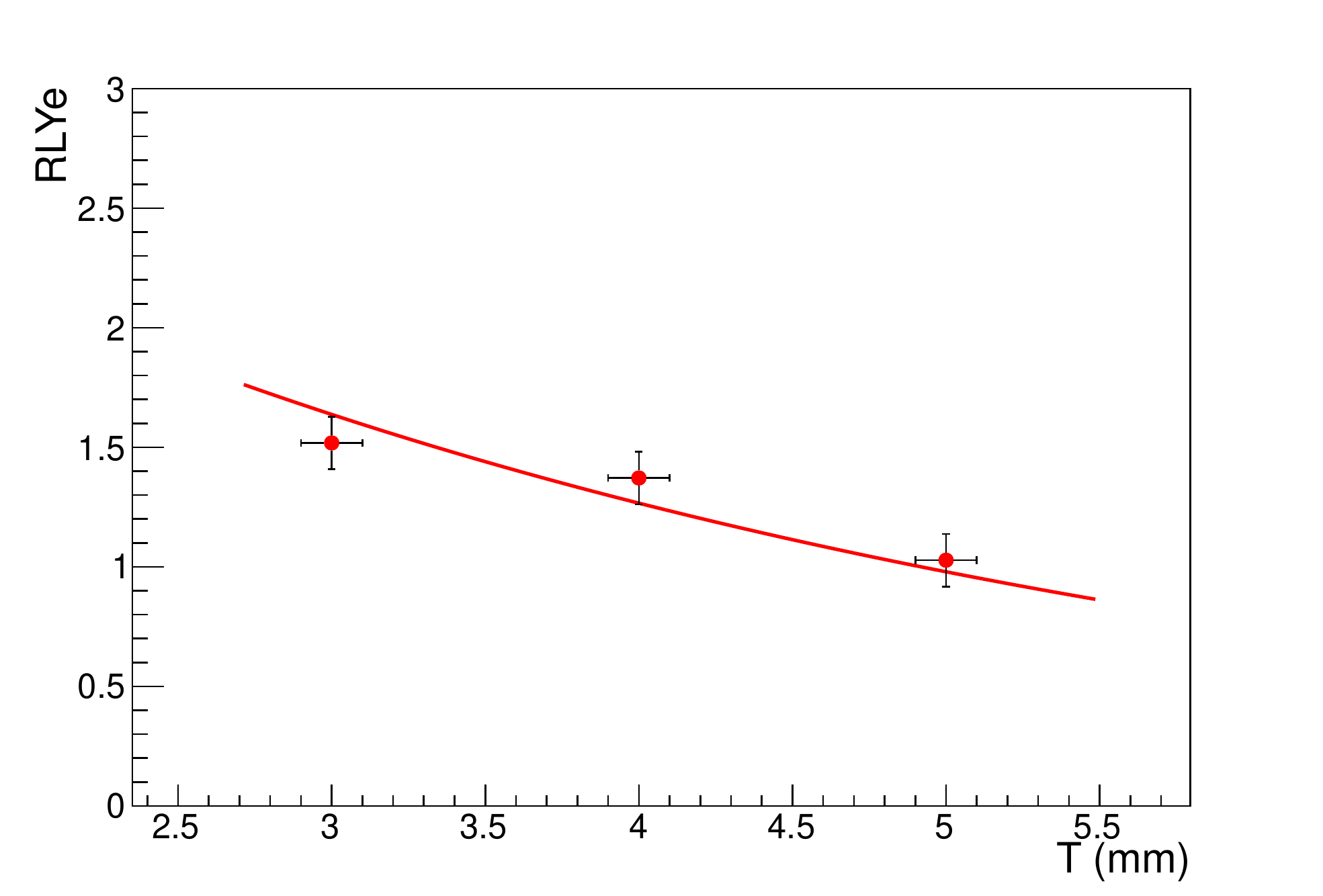}
\caption{Dependence  on the p-terphenyle sample thickness of the mean electron signal (top,  in red and EJ-200 in blue) and the  ratio between the yield in p-terphenyl  and EJ-200  (bottom). The superimposed fit is described in the text.}
\label{fig:LY}
\end{center}
\end{figure}

Fig.~\ref{fig:LY} top shows the electron detected signal as a function of the scintillator thickness ($T$), for the analyzed samples.
% A clear decrease of collected charge with increasing thickness is shown for p-terphenyl, while no appreciable effect is present for the EJ-200 samples. This is due to the short $\lambda$ for the p-terphenyl ($4.7\ \milli \meter$), and the long ($\sim 4\ \meter$) attenuation length for the ordinary plastic scintillators.
%In order to extrapolate the measured LY at zero thickness 
%In order to refer the LY to a scintillator with known yield, we have normalized it to the EJ-200 measured values (see Fig.~\ref{fig:LY} on the right).

By fitting  the distribution of the mean electron signals ($\mu$) as a function of the thickness $T$ with an exponential function ($y=LY_e\times e^{-x/{\lambda_e}}$), we measure %the parameter $LY_e=\pm $, and 
the Light Attenuation Length for $\sim 640\ \kilo \electronvolt$ electrons 

\begin{equation}
\label{eq:lambda_e}
\lambda_e=3.89\pm 0.15\ \milli \meter .
\end{equation} 

This value is in agreement with the one evaluated with the $\alpha$ source ($\lambda$), within 2.5 standard deviations. This indicates that the effect of the penetration spectrum of the $^{137}$Cs electrons is small. 
 %This effect is also confirmed by the simulation. 
Quantitatively, the ratio between the effective attenuation length estimated with electrons and the attenuation length measured with $\alpha$ particles is $R_\lambda =0.91\pm 0.04$ in data. It is found to be identical in MC.

Comparing the measured signals from electrons and $\alpha$ particles, as described in Sec.~\ref{LAL}, we can extract the $\alpha$ quenching. %Without normalizing to the Anthracene, the zero-thickness fit of the electron curve is $LY_e=YYYY$, while the corresponding value for the $\alpha$ was $LY_\alpha=YYYY$. 
Accounting for the difference in energy between the electron ($E_e=629\ \kilo \electronvolt$ on average) and the $\alpha$ lines ($E_\alpha=5480\ \kilo \electronvolt$ on average), the quenching factor $Q_\alpha$ can be estimated as:
\begin{equation}
Q_\alpha=R_{HV}\times\frac{LY_\alpha}{LY_e}\frac{E_e}{E_\alpha}
\label{eq:meanquench}
\end{equation}
with $LY_e=398\pm 22$ obtained from an exponential fit to Fig.~\ref{fig:LY},  $LY_\alpha=160\pm 4$ from the exponential fit to Fig.~\ref{fig:FITalpha}, both in units of PMT collected charge,  and $R_{HV}$ accounting for the different PMT High Voltage bias used with $\alpha$ particles ($HV_\alpha=1100\ \volt$) and electron source  ($HV_e=1300\ \volt$). 

To estimate the $R_{HV}$ correction, we measured the PMT gain curve as a function of the PMT High Voltage (HV) using a $3\times3\times5 \ \centi\meter^3$ LYSO scintillating crystal coupled to the PMT. A $^{137}$Cs source was used to excite the scintillation emission from the LYSO. The photo-peak from $662\ \kilo \electronvolt$ photons, visible because of the LYSO high $Z$, was used as a reference light source for the photoelectron emission by the PMT photocathode, to measure the charge collected at the PMT anode as a function of the HV. Measuring the peak position ($q$) and its width ($\sigma(q)$), the PMT gain was estimated as 
\begin{equation}
G=\frac{\sigma(q)^2}{e q}
\end{equation}
with $e$ the electron electric charge.

\begin{figure}[bth!]
\begin{center}
\includegraphics [width = 0.45  \textwidth] {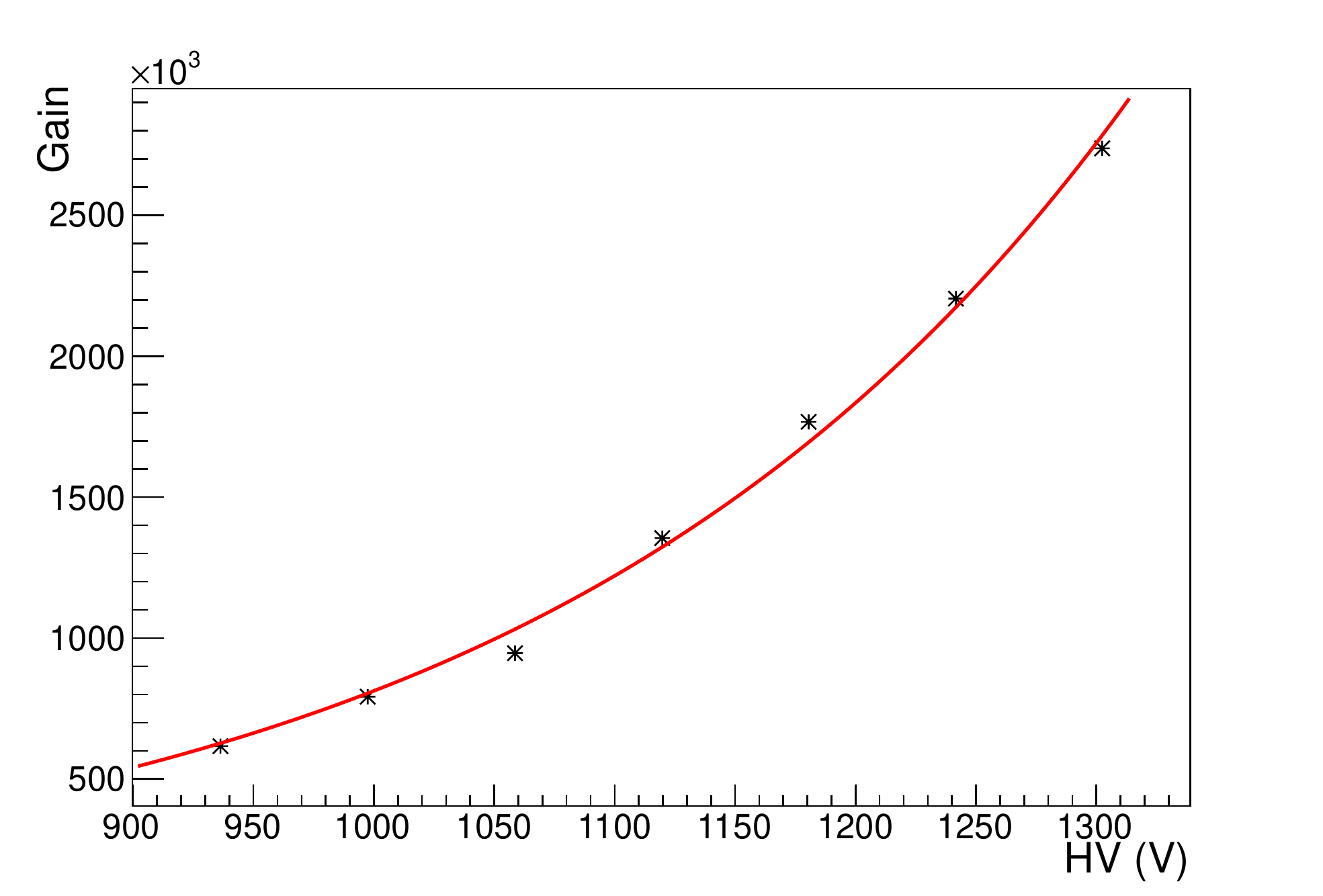}
\caption{Measured PMT gain as a function of the bias HV. The superimposed fit is described in the text.
\label{fig:gain} }
\end{center}
\end{figure}

Fig.\ref{fig:gain} shows the measured PMT gain as a function of the HV, and the power law curve ($G=p_0\times HV^{p_1}$) used to fit its distribution.  The correction factor is consequently estimated to be $R_{HV}=2.26\pm 0.04$, thus yielding the $\alpha$ quenching factor to be
\begin{equation}
Q_\alpha=(10.7\pm 0.6)\%.
\end{equation}

Finally, to measure the relative $LY$ of p-terphenyl, we have compared the signal released by the electrons in p-terphenyl to the one released in the EJ-200. In order to disentangle the effect of the light attenuation, we compared the LY for several values of $T$: $3$, $4$, and $5 \ \milli \meter$, combination of  EJ-200 samples being sensitive to optical coupling effects. Fig.~\ref{fig:LY} bottom shows the ratio between the signals detected with p-terphenyl and EJ-200 as a function of $T$. Since the EJ-200 shows no dependence on $T$, given its large attenuation length, we can fit the LY ratio with an exponential function ($RLY_e \times e^{-T/\lambda}$) with $\lambda$ fixed to the measured one (Eq.~\ref{eq:lambda_e}). 
Since, considering exclusively the statistical error, the $\chi^2$ of the fit is significantly above the number of degrees of freedom ($N_{DOF}$), we assign to the experimental points constant  systematic  errors to impose $\chi^2=N_{DOF}$ thus accounting for uncorrelated systematic errors.
As a result we obtain a zero-thickness $LY$ for p-terphenyl relative to EJ-200:
\begin{equation}
RLY_e=3.5\pm 0.2.
 \end{equation} 
This value can be regarded as the scintillation produced by this material in the approximation of infinite $\lambda$ and under the light collection conditions of our setup. For a typical LY of  $\simeq$10000 photons/MeV of the EJ-200, this would correspond to $LY\simeq35000$ photons/MeV for the p-terphenyl samples.

\section{Photons}
\label{phot}
\begin{figure*}[bth]
\begin{center}
\includegraphics [width = 0.45  \textwidth] {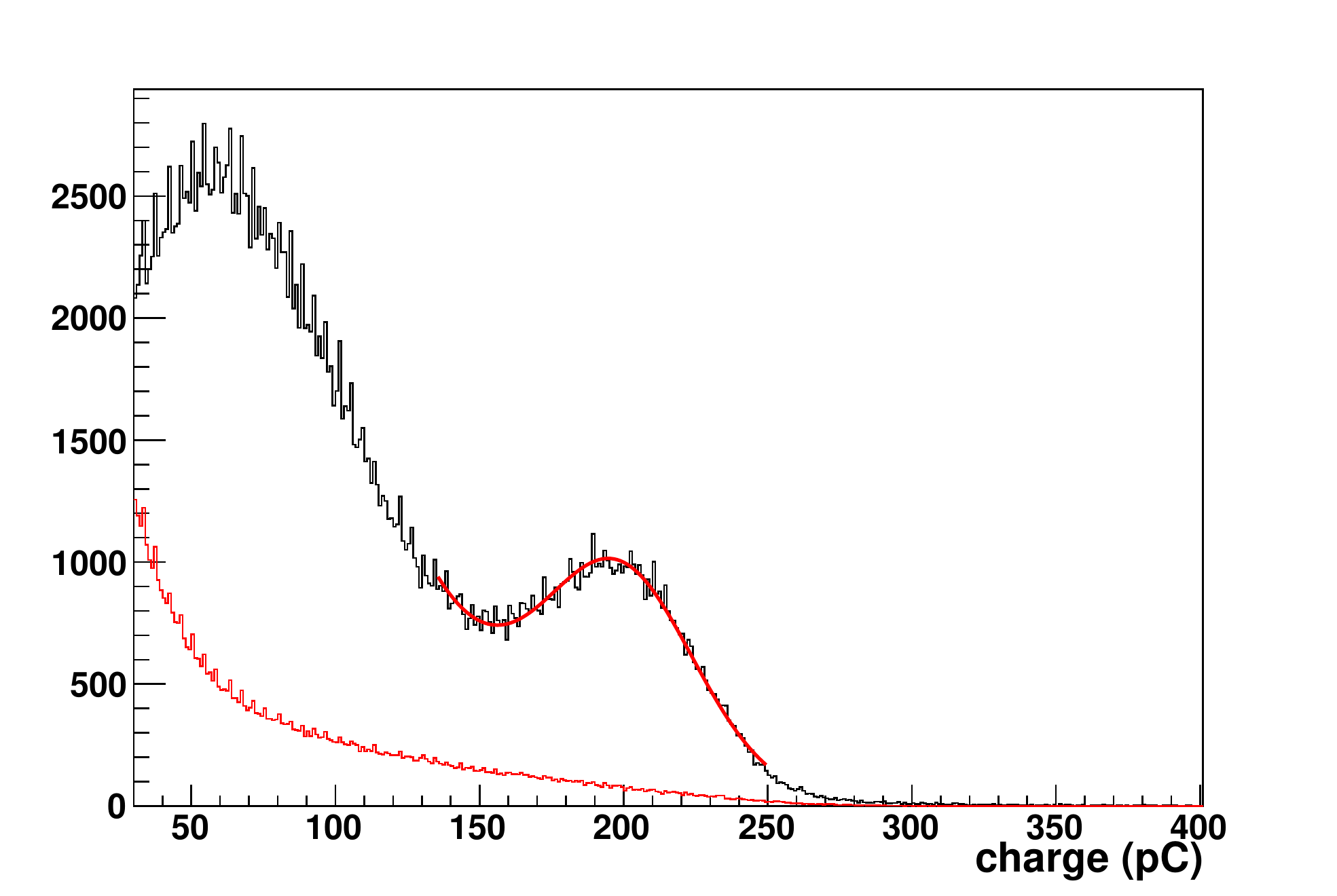}
\includegraphics [width = 0.45  \textwidth] {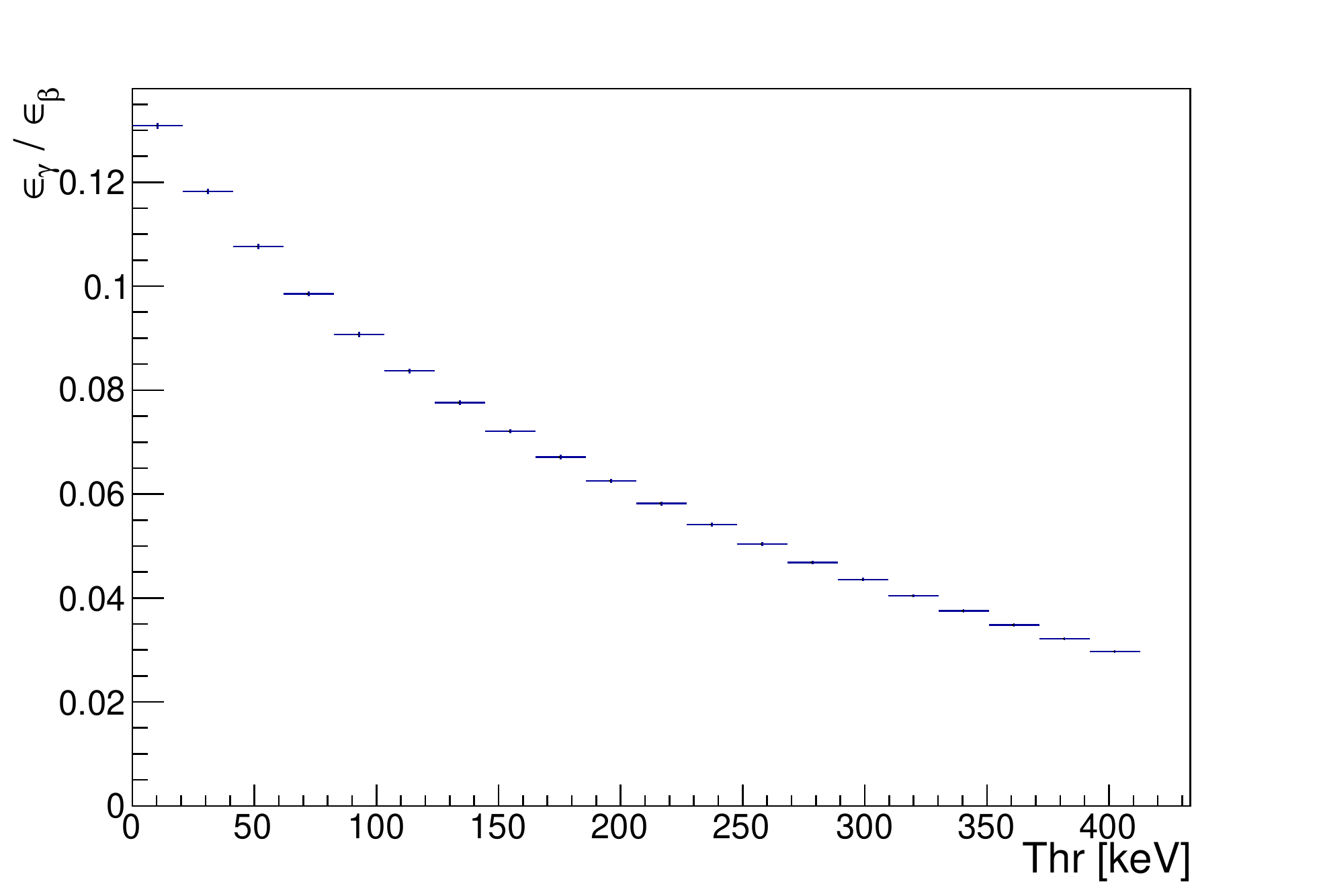}
\caption{Left: $^{137}$Cs spectra acquired with (red, lower histogram) and without (black, higher histogram) an aluminum layer in front of the detector. The fit superimposed to the latter spectrum allows to quantify the internal capture electron component. Right: estimated ratio between the efficiency to detect a $\sim660$ keV photon and  electron as a function of the threshold.}
\label{fig:gamma}
\end{center}
\end{figure*}

The small photon cross section of p-terphenyl, due to its low Z, can be exploited to detect electrons in environments with significant $\gamma$ pollution.
In order to quantify the suppression factor of photon detection with respect to electrons, we have used the decay properties of $^{137}$Cs. As detailed in Sec.~\ref{LY2},  a $662 \ \kilo \electronvolt$ photon is produced in $F_\gamma=78$\% of the cases and an almost monochromatic electron from internal capture in $F_e=8.7$\% of the cases (actually, two lines at $624$ and $656\ \kilo \electronvolt$).
While the electron lines appear as a single gaussian distribution in our detector, photons' energy spectrum is continuously decreasing with the increasing energy release.
 In order to distinguish the photon spectrum in the $^{137}$Cs emission, we have taken data interposing a $2 \ \milli \meter$ aluminum layer between the source and the detector. This layer absorbs the electron component of the $^{137}$Cs emission with no appreciable effect on the photon energy spectrum. The spectra obtained with and without aluminum are shown in Fig.~\ref{fig:gamma}.
The relative efficiency to detect a $\sim660 \ \kilo \electronvolt$ photon with respect to a $\sim 640\ \kilo \electronvolt$ electron ($R_\gamma$) is estimated with the following technique: the number of observed internal conversion electrons ($N_e$) is extracted from a gaussian fit to the energy spectrum obtained without aluminum shield; the number of photons which deposited an energy above a given threshold ($E_{thr}$, calibrated using the electron spectrum) is estimated by counting the selected events in the spectrum obtained with the aluminum layer ($N_\gamma(E_{thr})$).

 $R_\gamma$ as a function of the threshold is estimated as
\begin{equation}
R_\gamma(E_{thr})= \frac{N_\gamma(E_{thr})}{N_e} \frac{t_e}{t_\gamma} \frac{F_e}{F_\gamma}/C_{geom}, 
\end{equation}
with $t_\gamma$ and $t_e$ the elapsed times during the two acquisitions, $F_\gamma$ and $F_e$ the fractions of photons and electrons emitted by the $^{137}$Cs source as described above, and $C_{geom}$ the correction to the geometrical acceptance due to the aluminum layer interposition, estimated by MC (Sec.~\ref{MC}).
The assumptions made to obtain this result (mostly, the complete removal of the electron component with the aluminum and the lack of impact of the aluminum on the photon spectrum) were verified by means of the MC. We found the estimate to be correct at the 10\% level. 
The results are shown in Fig.~\ref{fig:gamma} on the right, the $\gamma$ suppression factor $R_\gamma$ ranging between 11\% for a $50 \ \kilo \electronvolt$ threshold to 3\% for a $300 \ \kilo \electronvolt$ threshold.

%%%%%%%%%%%%%%%%%%%%%%%%%%%%%%%%%%%%%%%%%%%%%%%%%%%%%%%%%%%%%%%%%%%%%%%%%%%%%%
\section{Simulation}
\label{MC}
To support the observations and allow the optimization of a possible detector, a simulation has been performed with FLUKA Monte Carlo (MC) software release 2011.2~\cite{FLUKA,fluka2}. The full detector geometry was included and the detailed optical simulation was enabled in order to include the generation of optical photons and their propagation, including attenuation effects, to the surface of the PMT. The simulations included in this paper are made with the following parameters:
\begin{itemize}
\item energy of the optical photon $E_{op}=2.96\ \electronvolt$;
%\item attenuation length $\lambda=4.73$\ \milli \meter\  as estimated in Sec.~\ref{LAL}; 
\item attenuation length $\lambda$; 
\item fraction of energy deposited in the case of $\alpha$ particles $f_\alpha$, which includes the $Q_\alpha$ quenching factor (Sec.~\ref{LY2});
\item fraction of electron energy deposited in the p-terphenyl that is converted into optical photons $f_e$.
\end{itemize}

\subsubsection*{Fraction of $\alpha$ deposited energy $f_\alpha$}
The $f_\alpha$ parameter in the simulation has been adjusted in order to match the experimental energy spectrum obtained with $\alpha$ particles. The optimal data-MC match, albeit not completely reproducing the low-energy tail, is found for $f_\alpha=0.01$, and the agreement is shown in Figs.~\ref{fig:alphaspectrum}. This estimated fraction includes the quenching of the $\alpha$ particles, $Q_{\alpha}$ as estimated in Sec.~\ref{LY2}.

\subsubsection*{Fraction of electron deposited energy $f_e$}
The value of fraction of electron energy deposited in the p-terphenyl, converted into optical photons, to be used in the simulation has been estimated using the $f_\alpha$ parameter and the measured $Q_\alpha$ quenching factor: 
$f_e = f_\alpha / Q_\alpha=0.1$. This implies that, when a 1 MeV electron is absorbed by the p-terphenyl, 100 keV of energy are converted into scintillation light. Assuming an average energy of the optical photons emitted by the p-terphenyl to be 2.96 eV, the corresponding light yield would be about 33,000 photons/MeV: this is compatible with the estimate obtained from the comparison with the EJ200 scintillator (Sec.~\ref{LY2}).

\subsubsection*{Light path length correction}
As explained in Sec.~\ref{LAL}, the scintillation light path from $\alpha$ particles includes the scintillator sample thickness $T$ plus an additional contribution due to the angular spread of optical photons. In order to account for this contribution, measurements with several thicknesses were simulated 
assuming $\lambda=5\ \milli \meter$ and performing the same analysis as on data. The dependence of the LY on the thickness is exponential (as in the data) but with an effective attenuation length of $4.5\ \milli \meter$. From this observation we introduced  a geometrical correction factor to translate the measured light yield $LY_m(T)$ into the true one $LY_t(T)=LY_m(T)\exp^{T/T_{geom}}$ and estimated $T_{geom}=45\ \milli \meter$.

\subsubsection*{Photon source geometrical acceptance}
The interposition of the aluminum layer changes the geometrical configuration of the source-detector system. We simulated the two different configurations in order to estimate the relative photon acceptance $C_{geom}$.

%%%%%%%%%%%%%%%%%%%%%%%%%%%%%%%%%%%%%%%%%%%%%%%%%%%%%%%%%%%%%%%%%%%%%%%%%%%%%%

%%%%%%%%%%%%%%%%%%%%%%%%%%%%%%%%%%%%%%%%%%%%%%%%%%%%%%%%%%%%%%%%%%%%%%%%%%%%%%
\section{Conclusions}
\label{conclu}
So far, p-terphenyl has not been used as main component of particle detectors due to its short Light Attenuation Length. As a matter of fact this property could be exploited when detecting low-energy charged particles in small areas, providing both energy and position measurement. To this extent p-terphenyl optical and scintillation properties need to be known.
 
A detailed study of p-terphenyl doped with 0.1\% diphenylbutadiene has been performed, in view of possible applications to radiation detectors. 
In particular, we investigated their response to $\alpha$ particles from $^{141}$Am decays, and  electrons and photons from $^{137}$Cs decays.

Optical and scintillation properties of these samples have been measured for the first time, in particular the Light Attenuation Length $\lambda=4.73\pm 0.06 \ \milli \meter$ and the Light Yield $RLY=3.5\pm 0.2$ relative to EJ-200 plastic scintillator. 

The $^{141}$Am source data show that using  p-terphenyl doped with 0.1\% diphenylbutadiene, the fraction of energy of the $\alpha$ particles converted into optical photons is reduced by a quenching factor $Q_\alpha=(10.7\pm0.6)\%$.  

Finally the sensitivity to photons has been studied to investigate electrons detection in highly $\gamma$-polluted environments. Photon suppression ranging between 3\% , at 300 keV threshold, and 11\% at 50 keV threshold has been observed.

A simulation based on FLUKA Monte Carlo has been developed exploiting these results, and has been used to properly account for acceptance and photons' path length corrections on data.  

%%%%%%%%%%%%%%%%%%%%%%%%%%%%%%%%%%%%%%%%%%%%%%%%%%%%%%%%%%%%%%%%%%%%%%%%%%%%%%

%%%%%%%%%%%%%%%%%%%%%%%%%%%%%%%%%%%%%%%%%%%%%%%%%%%%%%%%%%%%%%%%%%%%%%%%%%%%%%

%
%

%%%%%%%%%%%%%%%%%%%%%%%%%%%%%%%%%%%%%%%%%%%%%%%%%%%%%%%%%%%%%%%%%%%%%%%%%%%%%%
\end{document}